\documentclass[%
reprint,
prd,
superscriptaddress,nofootinbib
frontmatterverbose, 
 amsmath,amssymb,
 aps,
 showpacs,
floatfix,
]{revtex4-1}

\usepackage{graphicx}
\usepackage{dcolumn}
\usepackage{bm}
\usepackage{graphicx}
\usepackage{amsmath}
\usepackage{amssymb}
\usepackage{physics}
\usepackage{esvect}
\usepackage{cancel}
\usepackage{bbold}

\usepackage{geometry}
\usepackage{verbatim}
\usepackage{amsmath}
\usepackage{amssymb}
\usepackage{color}
\usepackage{soul}

\usepackage{tikz}
\usetikzlibrary{shapes.misc}
\usetikzlibrary{decorations.markings}

\usepackage[unicode=true,pdfusetitle,
 bookmarks=true,bookmarksnumbered=false,bookmarksopen=false,
 breaklinks=true,
 colorlinks=true,citecolor=steelblue]
 {hyperref}

\usepackage{url}

\makeatletter

\providecommand{\LyX}{\texorpdfstring%
  {L\kern-.1667em\lower.25em\hbox{Y}\kern-.125emX\@}
  {LyX}}

\makeatother

\definecolor{airforceblue}{rgb}{0.36, 0.54, 0.66}
\definecolor{steelblue}{rgb}{0.27, 0.51, 0.71}
\definecolor{amber}{rgb}{1.0, 0.49, 0.0}

\begin{document}


\title{Casimir-Polder interactions with massive photons: \\implications for BSM physics}

\author{\textsc{L.~Mattioli}}

\affiliation{Center for Theoretical Physics, Polish Academy of Sciences, Al.~Lotnik\'{o}w 32/46, 02-668 Warsaw, Poland}

\author{\textsc{A.~M.~Frassino}}
\affiliation{Departament de F{\'\i}sica Qu\`antica i Astrof\'{\i}sica, Institut de
Ci\`encies del Cosmos,\\ Universitat de
Barcelona, Mart\'{\i} i Franqu\`es 1, E-08028 Barcelona, Spain}

\author{\textsc{O.~Panella}} 
\email[{\bf Corresponding Author,}\\ Email: ]{orlando.panella@pg.infn.it }
\affiliation{Istituto Nazionale di Fisica Nucleare, Sezione di Perugia, Via A.~Pascoli, I-06123 Perugia, Italy}

\date{\today}

\begin{abstract}
We present the derivation of the Casimir-Polder interactions mediated by a massive photon between two neutral systems described in terms of their atomic polarizability tensors. 
We find a compact expression for the leading term at large distances between the two systems. Our result reduces, in the mass-less photon limit, to the standard Casimir-Polder. We discuss implications of our findings with respect to recent scenarios of physics beyond the standard model such as universal extra dimensions, Randall-Sundrum and scale-invariant models. For each model we compute the correction to the Casimir-Polder interaction in terms of the free parameters.
\end{abstract}

\maketitle
%
%
\section{\label{sec:introduction} Introduction}
The Casimir effect is the famous and fascinating quantum field theory phenomenon  whereby two  parallel plates (perfect conductors) separated a distance $a$ in vacuum attract each other with  interaction energy (Casimir energy) 
\begin{equation}
\label{CasimirEnergy}
{\cal E}^C = - \frac{\pi^2}{720} \frac{\hbar c}{a^3}
\end{equation}
and it is today widely interpreted as arising from the structure of the quantum vacuum~\cite{Milton_Book,Jaffe:2005vp,Graham:2002xq}. The interaction energy (and thus the  force) between the plates is due to the difference between  the vacuum energy of the electromagnetic field  without and with the plates (i.e. without and with geometrical boundary conditions). Such vacu\-um energies, though infinite by themselves, turn out to differ by a finite  amount which originates the measurable Casimir energy.
In typical Casimir effect experiments~\cite{Bordag:2009zzd} it is the Casimir force ${\cal P}^C = -\partial {\cal E}^C /\partial a $ that is actually measured.

It is interesting to note that historically H.~B.~Casimir computed initially~\cite{PhysRev.73.360} the interaction energy between two neutral systems (atoms or molecules) at distance $r$ from each other and characterized by static polarizabilities $\alpha_i(0), (i=1,2)$:
\begin{equation}\label{Massless Casimir - Polder}
 U(r) = - \frac{1}{{ {\left( 4 \pi \right)} }^3} \frac{23\, {\alpha}_1 {\left( 0 \right)} {\alpha}_2 {\left( 0 \right)} }{r^7}\, ,
\end{equation}
by starting with the usual van der Waals-London forces and correcting {it} for retardation effects. This was a standard second order  perturbation theory calculation in quantum mechanics.
Afterwards, apparently as a result of a conversation with Bohr~\cite{Milton:2011aa}, H.~B.~Casimir was able to show that the same result in Eq.~\eqref{Massless Casimir - Polder} could be derived  ``\emph{studying by means of  classical electrodynamics the change of the electromagnetic zero point energy}''~\cite{1949JCP....46..407C}.
Only later~\cite{Casimir:1948dh} he applied the same method of the vacu\-um fluctuations to derive the interaction energy between two perfectly conducting plates as in Eq.~\eqref{CasimirEnergy} which has become known as the Casimir effect.
It is one of the most celebrated mechanical effects of vacuum fluctuations~\cite{Resource}.

 From the experimental point of view, the first convincing measurement of the Casimir effect appeared only in 1997 when it was measured \cite{Lamoreaux:1997aa} in the range 0.6 to 6 $\mu$m for the configuration of a plane and a sphere whose force, when the distance is small compared to the radius of the sphere, can be deduced from that for parallel plates by using the proximity force approximation. Subsequently,  exact and reliable numerical calculations triggered more refined observations~\cite{Mohideen:1998aa,Decca:2005aa,Decca:2007aa,Decca:2007ab}. For a recent review see~\cite{Lamoreaux:2011aa}. It should be remarked that essentially all previously mentioned  measurements refer to the plane-sphere configuration. Indeed the measurement of the Casimir force between conductor plates is plagued with difficulties in maintaining the parallelism and by overwhelming electrostatic forces. The first precise measurement of the Casimir effect carried out in the configuration of the parallel plates was reported in \cite{Bressi:2002aa}.  The first conclusive measurements of the Casimir-Polder interactions measured the force between an atom and a pair of plates in a wedge configuration~\cite{Sukenik:1993aa}.

 It should also be  remarked that the Casimir energy and/or force in the geometry of the parallel conductors plates can be obtained via a pairwise integration of the Casimir-Polder interactions of the atoms and molecules making up the plates~\cite{Milton_Book,Bordag:2009zzd}. 

More recently,  the Casimir effect has attracted attention even in the field of superconductors. Indeed it is known that many types of superconducting detectors naturally form Casimir cavities (Superconducting Tunnel Junctions, and  Transition Edge Sensor geometries) for which Casimir forces could be relevant. In  these circumstances~\cite{Superconductors}, since gauge invariance is broken in a superconductor,  it is clear that  the understanding of the Casimir effect with a massive vector field (massive photon) may become relevant.   The Casimir force between two conducting lines (condensed vortices) for a massive scalar  field  has been investigated in~\cite{Vortices}.
 In connection with these aspects,  a pioneering  work to address the Casimir effect between perfect conductor plates with a massive photon is that of Barton and Dombey~\cite{Barton:1984kx,Barton:1985aa} where Proca electromagnetism~\cite{Proca_Stueckelberg,Photon,Photon_Graviton} is used to discuss in detail the mass dependence  in the limit of small photon mass. The Casimir effect with a massive vector field was then generalized to other geometries (spherical concentric)~\cite{Teo:2011aa} and the case of real metal plates (two  parallel dielectric planes)~\cite{Teo:2010aa,Teo:2012aa} including also a discussion of temperature effects.

%

In the literature, several beyond the Standard Model (BSM) scenarios have been already investigated in connection with the Casimir effect in recent years. For instance  within models with a generalized uncertainty principle (GUP) computing the first order correction in terms of the minimal length ($\hbar\sqrt{\beta}$)~\cite{Frassino:2011aa,Blasone:2019aa}. Other works include investigating the Casimir effect within Randall-Sundrum models \cite{Frank:2007ab,Frank:2008aa}, non commutative Randall-Sundrum models~\cite{Teo_5},  compactified universal extra dimensions (UED)~\cite{Poppenhaeger:2003es} and a scale invariant theory (unparticles)~\cite{Frassino:2017aa}. See also \cite{Frassino:2019aa} for an alternative approach  to the unparticle Casimir effect, based on the extended problem of Caffarelli and Silvestre~\cite{Caffarelli:2007aa}, of the quantization of the unparticle action of a scalar field with scaling dimension $d_{\cal U}$. The Casimir effect has also been investigated  in extended theories of gravity~\cite{Buoninfante:2019aa,Lambiase:2017aa} and Post-Newtonian gravity with Lorentz-violation~\cite{Blasone:2018nfy}. 
The Casimir force for parallel plates in  the spacetime  
with one extra space-like dimension is computed in terms of the decomposition into a Kaluza-Klein (KK) tower of massive vector fields in \cite{Teo:2010ab,Edery:2008kd}.%

On the other hand, one of the present authors investigated the effects of theories with a minimal length in  Casimir-Polder interactions~\cite{Panella:2007kd}.
To the best of our knowledge, however, the implications of theories characterized by a spectrum of massive particles have not yet been addressed in the realm of Casimir-Polder interactions. This work aims therefore at bridging such a gap by studying the Casimir-Polder interaction with a massive vector field (massive photon) and then applying the results to various BSM scenarios such as universal extra dimensions (UED), Randall-Sundrum models (RS) and scale invariant theories (unparticles).

The organization of the paper is as follows: in Sec.~\ref{sec:CPinteractions} we
shall review  the derivation of the Casimir-Polder interactions: after reviewing the Casimir-Polder within ordinary quantum electrodynamics (QED) we provide the computation of the Casimir-Polder interaction mediated by a massive vector field (massive photon). In Sec.~\ref{sec:bsm} we discuss the Casimir-Polder energy arising within some interesting scenarios of physics beyond the standard model such as universal extra dimensions, Randall-Sundrum models and  a model with scale invariance, i.e. when the interaction is mediated by an unparticle vector field.
Finally Sec.~\ref{sec:conclusions} is
dedicated to our conclusions.

\section{The Casimir-Polder interaction with a massive photon \label{sec:CPinteractions}}

In this section, we offer a complete derivation of the Casimir - Polder Force both for the electromagnetic field - i.e. the standard Casimir-Polder Force - and for a massive electromagnetic field.

We provide  a unified approach, where the first part  of the computation  is valid both in the standard QED (massless) and in the massive case while in the second part we consider separately the massless and the massive results and in the end we check that the massless limit of the latter is equal to the former. 

We introduce first the free electromagnetic field and then the interaction between the electromagnetic field and two non - polar molecules.

The Lagrangian density of a free vector field $A^{\mu}$ of mass $\mu$ is the Proca Lagrangian density 
\begin{equation}
\begin{split}
{\cal L} = & - \frac{1}{4} F_{\mu \nu} F^{\mu \nu} + \frac{1}{2} {\mu}^2 A_{\mu} A^{\mu} = \\
& \frac{1}{2} A_{\mu} {\left[ {\left( \Box + {\mu}^2 \right)} {g^{\mu}}_{\nu} - {\partial}^{\mu} {\partial}_{\nu} \right]} A^{\nu} - \frac{1}{2} {\partial}_{\mu} {\left( F^{\mu \nu} A_{\nu} \right)}
\end{split}
\end{equation}
where
\begin{equation}
F^{\mu \nu} = {\partial}^{\mu} A^{\nu} - {\partial}^{\nu} A^{\mu}\, ,
\end{equation}
and the conjugate momenta of the field $A^\mu(x)$ is:
\begin{equation}
    \Pi^\mu(x) = F^{\mu0}(x).
\end{equation} 
Then quantization of the Proca theory is carried out by imposing canonical equal time commutation relations:
\begin{subequations}
\label{ETCR}
\begin{align}
\delta(t_1-t_2)\left[A^\mu(x_1),A^\nu(x_2)\right]\,=\,&0;\label{ETCRa}\\
\delta(t_1-t_2)\left[A^\mu(x_1),\Pi^{\nu}(x_2)\right]\,=\,& i\, g^{\mu\nu} \delta^4(x_1-x_2).
\label{ETCRb}
\end{align}
\end{subequations}
Since the Lagrangian density is defined up to a divergence it can be rewritten as 
\begin{equation}
\label{finalL}
{\cal L} = \frac{1}{2} A_{\mu} {\left[ {\left( \Box + {\mu}^2 \right)} {g^{\mu}}_{\nu} - {\partial}^{\mu} {\partial}_{\nu} \right]} A^{\nu}
\end{equation}
The presence of the term in the derivatives of the second order in Eq.~\eqref{finalL} modifies~\cite{Riahi:1972aa} the standard Euler\-Lagrange equations as:
\begin{equation}
\frac{\partial {\cal L}}{\partial A_{\mu}} - {\partial}_{\alpha} \frac{\partial {\cal L}}{\partial {\partial}_{\alpha} A_{\mu}} + {\partial}_{\alpha} {\partial}_{\beta} \frac{\partial {\cal L}}{\partial {\partial}_{\alpha} {\partial}_{\beta} A_{\mu}} = 0
\end{equation}
By substituting the Lagrangian density into the Euler-\-Lagrange equation we get the Proca equation \\
\begin{equation}
{\left( \Box + {\mu}^2 \right)} A^{\mu} - {\partial}^{\mu} {\partial}_{\nu} A^{\nu} = 0\,.
\end{equation}
In the massless case, the Proca equation becomes the Maxwell equation 
\begin{equation}
\Box A^{\mu} - {\partial}^{\mu} {\partial}_{\nu} A^{\nu} = 0\,.
\end{equation}
In the massive case, by taking its divergence the Proca equation can be rewritten as 
\begin{equation}
\begin{cases}
{\left( \Box + {\mu}^2 \right)} A^{\mu} = 0 \, ,\\
{\partial}_{\mu} A^{\mu} = 0\, ,
\end{cases}
\end{equation}
where the equations are formally equal to the Klein-Gordon equation and the Lorentz Gauge, respectively. The Feynman propagator in momentum space is obtained by inverting the Fourier-transformed differential operator contained in the Lagrangian density \cite[page 188]{Greiner} 
\begin{equation}
\begin{split}
{{ {\left( D^{-1} \right)} }^{\mu}}_{\nu} {\left( k \right)} = & - {\left( k^2 - {\mu}^2 \right)} {g^{\mu}}_{\nu} + k^{\mu} k_{\nu} = \\
& - {\left( k^2 - {\mu}^2 \right)} {\left( {g^{\mu}}_{\nu} - \frac{k^{\mu} k_{\nu}}{k^2} \right)} + {\mu}^2 \frac{k^{\mu} k_{\nu}}{k^2}
\end{split}
\end{equation}
Since the last expression is a spectral representation we get 
\begin{eqnarray}
{{ {\left[ f {\left( D^{-1} \right)} \right]} }^{\mu}}_{\nu} {\left( k \right)} &=& f {\left[ - {\left( k^2 - {\mu}^2 \right)} \right]} {\left( {g^{\mu}}_{\nu} - \frac{k^{\mu} k_{\nu}}{k^2} \right)} \nonumber \\ &&\phantom{xxxxxxxxxxx}  +\, f {\left( {\mu}^2 \right)} \frac{k^{\mu} k_{\nu}}{k^2}
\end{eqnarray}
where $f {\left( x \right)}$ is any function. 
For $f {\left( x \right)} = x^{-1}$ we get 
\begin{equation}
{{D}^{\mu}}_{\nu} {\left( k \right)} =  - \frac{{g^{\mu}}_{\nu} - \frac{k^{\mu} k_{\nu}}{k^2}}{k^2 - {\mu}^2} + \frac{\frac{k^{\mu} k_{\nu}}{k^2}}{{\mu}^2} =
 - \frac{{g^{\mu}}_{\nu} - \frac{k^{\mu} k_{\nu}}{{\mu}^2}}{k^2 - {\mu}^2}
\end{equation}
and the Feynman propagator in momentum space is therefore 
\begin{equation}
D^{\mu \nu} {\left( k \right)} = \frac{- g^{\mu \nu} + \frac{k^{\mu} k^{\nu}}{{\mu}^2}}{k^2 - {\mu}^2 + i 0^{^+}}\, ,
\end{equation}
where the pole is shifted as usual by adding a small negative imaginary part to the mass in order to satisfy the causality condition~\cite[page 188]{Greiner}. 
The Feynman propagator in position space is obtained by Fourier anti-transforming:
\begin{equation}
D^{\mu \nu} {\left( x \right)} = {\left( - g^{\mu \nu} - \frac{{\partial}^{\mu} {\partial}^{\nu}}{{\mu}^2} \right)}  \int  \frac{d^4 k}{(2 \pi)^4}\frac{e^{- i k x}}{k^2 - {\mu}^2 + i 0^{^+}} \, .
\end{equation}
The appearance of a divergent term as $\mu\to 0$ could lead  to the naive conclusion that it is not possible to recover standard quantum electrodynamics, by taking the mass-less limit of the Proca theory. However, gauge invariance will save the day. Computation of physical observables will involve gauge invariant quantities like for instance the correlation functions of the field strength tensor components (electric and/or magnetic fields). These correlations functions will have a finite $\mu \to 0$
 limit. 
Indeed by making use of the equal time commutation relations given in Eqs.~(\ref{ETCRa},\ref{ETCRb}) the following identity can be  proved: 
\begin{widetext}
\begin{subequations}\label{Braket Tensor}
\begin{align}
\begin{split}
\label{BraketTensorA}
\expval{T {\left[ F^{\alpha \beta} {\left( x_1 \right)} F^{\gamma 0} {\left( x_2 \right)} \right]} }{0} &= +{\left( {{\partial}_1}^{\alpha} {g^{\beta}}_{\mu} - {{\partial}_1}^{\beta} {g^{\alpha}}_{\mu} \right)} {\left( {{\partial}_2}^{\gamma} {g^{0}}_{\nu} - {{\partial}_2}^{0} {g^{\gamma}}_{\nu} \right)} \expval{T {\left[ A^{\mu} {\left( x_1 \right)} A^{\nu} {\left( x_2 \right)} \right]} }{0} \\ 
&\phantom{=}- i (g^\alpha_{\phantom{\alpha} 0} g^{\beta\gamma} -g^\beta_{\phantom{\beta}0}g^{\alpha\gamma}) \delta^4(x_1-x_2)
\end{split}\\
\begin{split}
\label{BraketTensorB}
 &= 
 +i {\left( {{\partial}_1}^{\alpha} {g^{\beta}}_{\mu} - {{\partial}_1}^{\beta} {g^{\alpha}}_{\mu} \right)} {\left( {{\partial}_2}^{\gamma} {g^{0}}_{\nu} - {{\partial}_2}^{0} {g^{\gamma}}_{\nu} \right)} D^{\mu \nu} {\left( x_1 - x_2 \right)}  \\ &\phantom{=} - i (g^\alpha_{\phantom{\alpha} 0} g^{\beta\gamma} -g^\beta_{\phantom{\beta}0}g^{\alpha\gamma}) \delta^4(x_1-x_2)\end{split}\\ 
 \begin{split}
 \label{BraketTensorC}
 &=  +i {\left( {\partial}^{\alpha} {g^{\beta}}_{\mu} - {\partial}^{\beta} {g^{\alpha}}_{\mu} \right)} {\left( - {\partial}^{\gamma} {g^{0}}_{\nu} + {\partial}^{0} {g^{\gamma}}_{\nu} \right)} {\left( - g^{\mu \nu} - \frac{{\partial}^{\mu} {\partial}^{\nu}}{{\mu}^2} \right)}  \int \frac{d^4 k}{(2\pi)^4} \frac{e^{- i k {\left( x_1 - x_2 \right)} }}{k^2 - {\mu}^2 + i 0^{^+}} \\ 
 &\phantom{=}  - i (g^\alpha_{\phantom{\alpha} 0} g^{\beta\gamma} -g^\beta_{\phantom{\beta}0}g^{\alpha\gamma}) \delta^4(x_1-x_2)\end{split} \\
 \begin{split}
 \label{BraketTensorD}&=
+i {\left( {\partial}^{\alpha} {g^{\beta}}_{\mu} - {\partial}^{\beta} {g^{\alpha}}_{\mu} \right)} {\left( {\partial}^{\gamma} g^{0 \mu} - {\partial}^{0} g^{\gamma \mu} \right)}  \int \frac{d^4 k}{(2\pi)^4} \frac{e^{- i k {\left( x_1 - x_2 \right)} }}{k^2 - {\mu}^2 + i 0^{^+}}  \\
&\phantom{=}- i (g^\alpha_{\phantom{\alpha} 0} g^{\beta\gamma} -g^\beta_{\phantom{\beta}0}g^{\alpha\gamma}) \delta^4(x_1-x_2)\,.
\end{split}
\end{align}
\end{subequations}
\end{widetext}
It can be  seen that the divergent terms cancel out each other in Eq.~(\ref{BraketTensorC}) and the above correlations admit indeed a well defined massless limit $\mu\to 0$. 

The interaction Hamiltonian between the electromagnetic field and the neutral systems (atomes/molecules) placed at the positions $\bm{r}_1$ and $\bm{r}_2$ (with $\bm{r}_2 \neq \bm{r}_1$) is the dipole interaction: 
\begin{equation}\label{Dipole Interaction}
H_{int} {\left( t \right)} = - \bm{E} {\left( t,{\bm{r}}_1 \right)} \cdot {\bm{d}}_{(1)} {\left( t \right)} - \bm{E} {\left( t,{\bm{r}}_2 \right)} \cdot {\bm{d}}_{(2)} {\left( t \right)}
\end{equation}
where $\bm{E} {\left( t,\bm{r} \right)}$ is the electric field operator, ${\bm{r}}_k$ is the position and ${\bm{d}}_{(k)}$ is the electric dipole moment operator of the $k^{th}$ atom/molecule $(k=1,2)$. 

Since the neutral atoms/molecules are assumed to be non-polar $\expval{{\bm{d}_{(k)}}  }{{\psi}} = 0$ the potential energy from second order perturbation theory~\cite[page 348]{Landau}
vanishes.
 
The first non vanishing contribution to the potential energy comes then from fourth order perturbation theory~\cite[page 348]{Landau} 
\begin{eqnarray}\label{Potential Energy}
U {\left( \mu;{\bm{r}}_1,{\bm{r}}_2 \right)} &=&  \frac{i}{2 t} \int d t_1 d t_2 d t_3 d t_4 \, \times
\nonumber \\&&  \expval{T {\left[ E^i {\left( t_1,{\bm{r}}_1 \right)} E^j {\left( t_2,{\bm{r}}_2 \right)} \right]} }{0} \times \nonumber \\
&&  \expval{T {\left[ {d_2}^j {\left( t_2 \right)} {d_2}^l {\left( t_4 \right)} \right]} }{{\psi}_2} \nonumber \times \\
&& \expval{T {\left[ E^l {\left( t_4,{\bm{r}}_2 \right)} E^k {\left( t_3,{\bm{r}}_1 \right)} \right]} }{0} \times\nonumber \\&&  \expval{T {\left[ {d_1}^k {\left( t_3 \right)} {d_1}^i {\left( t_1 \right)} \right]} }{{\psi}_1} 
\end{eqnarray}
which in the language of Feynman diagrams is represented by a loop diagram with a two photon exchange. 

When computing the correlation function $\langle 0|T \left[ E^i  {\left( t_1,{\bm{r}}_1 \right)}  E^j {\left( t_2,{\bm{r}}_2 \right)} \right]|0\rangle$ the commutation of the time derivatives which enter the electric fields $E^i$ with the chronological $T-$product introduces terms proportional to $\delta^4(x_1-x_2) =\delta(t_1-t_2)\, \delta^3(\bm{r}_1-\bm{r}_2)$, as can be seen from Eq.~\eqref{BraketTensorA}. Since we are interested in the interaction energy between the two systems located at $\bm{r}_1, \bm{r}_2$ with $\bm{r}_1 \ne \bm{r}_2$, clearly $\delta^3(\bm{r}_1-\bm{r}_2)=0$.   Such terms proportional to the Dirac distribution $\delta^4(x_1-x_2)$ can be safely ignored and we conclude that time derivatives can be taken out of the $T-$product like the corresponding spatial derivatives. Therefore  from Eq.~\eqref{BraketTensorD} we get:
\begin{eqnarray}\label{Braket Vector}
\langle 0|&&T \left[ E^i  {\left( t_1,{\bm{r}}_1 \right)}  E^j {\left( t_2,{\bm{r}}_2 \right)} \right]|0\rangle \nonumber \\ && =\expval{T {\left[ F^{i 0} {\left( t_1,{\bm{r}}_1 \right)} F^{j 0} {\left( t_2,{\bm{r}}_2 \right)} \right]} }{0} \nonumber\\
&&= i {\left( {\partial}^i {\partial}^j - {\partial}^0 {\partial}^0 {\delta}^{i j} \right)}\!\!  \int\limits_{- \infty}^{\infty}\!\! \frac{d\omega}{2 \pi}D {\left( \omega,{\bm{r}}_1\! -\! {\bm{r}}_2 \right)} e^{- i \omega {\left( t_1 - t_2 \right)} } \nonumber\\
&& = i  \int\limits_{- \infty}^{\infty} \frac{d\omega}{2 \pi}{\left( {\omega}^2 {\delta}^{i j} + {\partial}^i {\partial}^j \right)} D {\left( \omega,{\bm{r}}_1\! -\! {\bm{r}}_2 \right)} e^{- i \omega {\left( t_1 - t_2 \right)} } \nonumber \\
\end{eqnarray}
where we have defined the scalar function $D {\left( \omega,\bm{r} \right)}$ as:
\begin{equation}
\label{Dfunction}
D {\left( \omega,\bm{r} \right)} =  \int \frac{d^3 \bm{k}}{{ {\left( 2 \pi \right)} }^3}\, \frac{e^{i \bm{k} \cdot \bm{r}}}{k^2 - {\mu}^2 + i 0^{^+}} \, .
\end{equation}
It can be shown that \cite[page 351]{Landau}%
\begin{equation}\label{Braket Dipole}
\expval{T {\left[ {d_k}^i {\left( t_1 \right)} {d_k}^j {\left( t_2 \right)} \right]} }{{\psi}_k} = i  \int\limits_{- \infty}^{\infty}\frac{d \omega}{2 \pi} \alpha_k^{i j} {\left( \omega \right)} e^{- i \omega {\left( t_1 - t_2 \right)} } 
\end{equation}
where ${{\alpha}_k}^{i j}$ is the polarizability tensor of the $k^{th}$ molecule and ${{\alpha}_k}^{i j} {\left( - \omega \right)} = {{\alpha}_k}^{i j} {\left( \omega \right)}$. 
By substituting  Eq.~\eqref{Braket Vector} and \eqref{Braket Dipole} in Eq.~\eqref{Potential Energy} and recognizing the  Dirac delta functions we get: 
\begin{widetext}
\begin{equation}
U {\left( \mu{;}\,{\bm{r}}_1 {-} {\bm{r}}_2 \right)} {=} \frac{1}{4 \pi} i {\int_{{-} \infty}^{\infty}} {\left[ {\left( {\omega}^2 {\delta}^{i j} {+} {\partial}^i {\partial}^j \right)} D {\left( \omega{,}{\bm{r}}_1 {-} {\bm{r}}_2 \right)} \right]} {{\alpha}_2}^{j l} {\left( \omega \right)} {\left[ {\left( {\omega}^2 {\delta}^{l k} {+} {\partial}^l {\partial}^k \right)} D {\left( \omega{,}{\bm{r}}_2 {-} {\bm{r}}_1 \right)} \right]} {{\alpha}_1}^{k i} {\left( \omega \right)} d \omega
\end{equation}
We readily recognize a trace in the expression above 
\begin{equation}
U {\left( \mu;\,{\bm{r}}_1 - {\bm{r}}_2 \right)} = \frac{1}{4 \pi} i \int_{- \infty}^{\infty} \text{Tr}\,  {\left[ {\left( {\omega}^2\mathbb{1} + \mathbb{H} \right)} D {\left( \omega,{\bm{r}}_1 - {\bm{r}}_2 \right)} \right]} {\mathbb{a}}_2 {\left( \omega \right)} {\left[ {\left( {\omega}^2\mathbb{1} + \mathbb{H} \right)} D {\left( \omega,{\bm{r}}_2 - {\bm{r}}_1 \right)} \right]} {\mathbb{a}}_1 {\left( \omega \right)} d \omega
\end{equation}
\end{widetext}
where $\mathbb{H}=\partial^i\partial^j$ is the Hessian matrix, $\mathbb{1}=\delta^{ij}$ is the identity operator and $\mathbb{a}_k=\alpha_k^{ij} (k=1,2)$ are the polarization tensors of the two neutral systems.

By using the facts that $D {\left( - \omega,\bm{r} \right)} = D {\left( \omega,\bm{r} \right)}$ and ${\mathbb{a}}_k {\left( - \omega \right)} = {\mathbb{a}}_k {\left( \omega \right)}$ we get: 
\begin{eqnarray}
U\left( \mu;{\bm{r}}_1 - {\bm{r}}_2 \right) &=&   i \int_0^\infty \frac{d\omega}{2\pi}\, \text{Tr} \left[ \left( \omega^2 \mathbb{1} + \mathbb{H} \right) D \left( \omega,\bm{r}_1 - \bm{r}_2 \right) \right. \nonumber \\ &&\left. \times \mathbb{a}_2 \left( \omega \right)  \left(\omega^2 \mathbb{1}+ \mathbb{H} \right) D {\left( \omega,{\bm{r}}_2 - {\bm{r}}_1 \right)} \right] \mathbb{a}_1(\omega) \nonumber \\
\end{eqnarray}
By using the fact that $D {\left( \omega,\bm{r} \right)} = D {\left( \omega,{\left| \bm{r} \right|} \right)}$ and considering the case of isotropic molecules $\alpha_{k}^{ij} (\omega) = \alpha_{k}(\omega) \delta^{ij}$~\footnote{or by replacing the polarizabilities ${\mathbb{a}}_k$ with the mean polarizabilities $\left( \text{Tr} {\mathbb{a}}_k \right)/3$}  follows 
\begin{equation}\label{Isotropic}
U {\left( \mu;r \right)} =   i \int_0^{\infty} \frac{d\omega}{2 \pi}{\alpha}_1 {\left( \omega \right)} {\alpha}_2 {\left( \omega \right)} \, \text{Tr}\, { {\left[ {\left( {\omega}^2 \mathbb{1} + \mathbb{H} \right)} D {\left( \omega,r \right)} \right]} }^2 
\end{equation}
where $r = {\left| {\bm{r}}_1 - {\bm{r}}_2 \right|}$.
By expanding ${\alpha}_k {\left( \omega \right)}$ we can compute ${\alpha}_1 {\left( \omega \right)} {\alpha}_2 {\left( \omega \right)}$ as a product of two series (Cauchy product), and by using again  ${\alpha}_k {\left( - \omega \right)} = {\alpha}_k {\left( \omega \right)}$ we get the following series
\begin{eqnarray}\label{Series}
{\alpha}_1 {\left( \omega \right)} {\alpha}_2 {\left( \omega \right)} &=& \sum_{n=0}^{\infty} \frac{{\omega}^{2 n}}{ {\left( 2 n \right)} !} \times \nonumber \\ &&\sum_{k=0}^n \binom{2 n}{2 k} \frac{d^{2 k} {\alpha}_1}{d {\omega}^{2 k}} {\left( 0 \right)} \frac{d^{2(n-k)} {\alpha}_2}{d {\omega}^{2(n-k)}} {\left( 0 \right)},\nonumber \\
\end{eqnarray}
and by using the fact that the Laplacian is the trace of the Hessian matrix it follows:
\begin{eqnarray}
\label{traceHD}
\text{Tr} [ ( \omega^2\mathbb{1} &+& \mathbb{H} ) D (\omega r ) ]^2 = 3 {\omega}^4 D^2 {\left( \omega,r \right)} \nonumber \\ && + 2 {\omega}^2 D {\left( \omega,r \right)} {\nabla}^2 D {\left( \omega,r \right)} + \text{Tr} {\left\{ { {\left[ \mathbb{H} D {\left( \omega,r \right)} \right]} }^2 \right\}}\nonumber \\
\end{eqnarray}
and in spherical coordinates:
\begin{equation}
\mathbb{H} D {\left( \omega r \right)} = \frac{1}{r} \frac{\partial D}{\partial r} {\left( \omega r \right)} {\left( \mathbb{1} - \frac{x x^T}{r^2} \right)} + \frac{{\partial}^2 D}{\partial r^2} {\left( \omega,r \right)} \frac{x x^T}{r^2}
\end{equation}
where $\mathbb{1}$ is the identity matrix, $x$ is the position column vector, ${x}^T$ denotes the transposed position vector and, of course, we have used only the radial term of the Hessian matrix. Since this is a spectral representation  one can write: 
\begin{eqnarray}
f {\left[ \mathbb{H} D {\left( \omega,r \right)} \right]} &=& f {\left[ \frac{1}{r} \frac{\partial D}{\partial r} {\left( \omega,r \right)} \right]} {\left( \mathbb{1} - \frac{x x^T}{r^2} \right)}\nonumber \\ && \phantom{xxxx} + f {\left[ \frac{{\partial}^2 D}{\partial r^2} {\left( \omega,r \right)} \right]} \frac{x x^T}{r^2}
\label{anyf}
\end{eqnarray}
where $f {\left( x \right)}$ is a function. 
And, since we need to compute Tr${\left[ \mathbb{H} D {\left( \omega,r \right)} \right]^2}$ in Eq.~\eqref{traceHD}, for $f {\left( x \right)} = x^2$,  the previous Eq.~\eqref{anyf} is:
\begin{eqnarray}
{ {\left[ \mathbb{H} D {\left( \omega,r \right)} \right]} }^2 &=& { {\left[ \frac{1}{r} \frac{\partial D}{\partial r} {\left( \omega,r \right)} \right]} }^2 {\left( \mathbb{1} - \frac{x x^T}{r^2} \right)} \nonumber \\ && \phantom{xxxx}+ { {\left[ \frac{{\partial}^2 D}{\partial r^2} {\left( \omega,r \right)} \right]} }^2 \frac{x x^T}{r^2}.
\end{eqnarray}
The computation of the trace
\begin{equation}
\text{Tr} { {\left[ \mathbb{H} D {\left( \omega,r \right)} \right]} }^2 = 2 { {\left[ \frac{1}{r} \frac{\partial D}{\partial r} {\left( \omega,r \right)} \right]} }^2 + { {\left[ \frac{{\partial}^2 D}{\partial r^2} {\left( \omega,r \right)} \right]} }^2
\end{equation}
finally gives:
\begin{eqnarray}\label{Trace}
\text{Tr} { {\left[ {\left( {\omega}^2\mathbb{1} {+} \mathbb{H} \right)} D {\left( \omega{,}r \right)} \right]} }^2 &=& \,3\, {\omega}^4 D^2 {\left( \omega{,}r \right)} \nonumber \\ &{+}&\, 2 {\omega}^2 D {\left( \omega{,}r \right)} \frac{1}{r^2} \frac{\partial}{\partial r} r^2 \frac{\partial D}{\partial r} {\left( \omega{,}r \right)} \nonumber \\ &{+}&\, 2 { {\left[ \frac{1}{r} \frac{\partial D}{\partial r} {\left( \omega{,}r \right)} \right]} }^2 {+} { {\left[ \frac{{\partial}^2 D}{\partial r^2} {\left( \omega{,}r \right)} \right]} }^2.\nonumber \\
\end{eqnarray}
By substituting Eq.~\eqref{Series} and Eq.~\eqref{Trace} in Eq.~\eqref{Isotropic} we obtain the explicit expression for the potential energy:
\begin{eqnarray}\label{Casimir - Polder Series}
 U \left( \mu;r \right) &=&\,  i 
 \sum_{n=0}^\infty \int_0^\infty  \frac{d\omega}{2\pi} \, \frac{{\omega}^{2 n}}{ {\left( 2 n \right)} !} \,\left\{ 3 \omega^4 D^2 \left( \omega{,}r \right)\phantom{\left[\frac{1}{r}\right]^2}\right. \nonumber \\  &&\,+ 2 \omega^2 D \left( \omega{,}r \right) \frac{1}{r^2} \frac{\partial}{\partial r} r^2 \frac{\partial D}{\partial r} \left( \omega{,}r \right)  \nonumber \\ &&{+} \left.  2 { {\left[ \frac{1}{r} \frac{\partial D}{\partial r} {\left( \omega{,}r \right)} \right]} }^2 +{ {\left[ \frac{{\partial}^2 D}{\partial r^2} {\left( \omega{,}r \right)} \right]} }^2 \right\} \times \nonumber\\
&& \sum_{k=0}^n \binom{2 n}{2 k} \frac{d^{2 k} {\alpha}_1}{d {\omega}^{2 k}} {\left( 0 \right)} \frac{d^{2 n - 2 k} {\alpha}_2}{d {\omega}^{2 n - 2 k}} {\left( 0 \right)}.
\end{eqnarray}
This will be the starting point for our analysis in the next subsections.
\subsection{Casimir-Polder interaction for massless photons}\label{submassless}
The scalar function $D(\omega,\bm{r})$ defined in Eq.~\eqref{Dfunction} in the massless photon case (when $\mu=0)$ is easily computed by using standard methods as:
\begin{equation}\label{Massless Propagator}
D {\left( \omega,r \right)} =   \int\!\!\frac{d^3 \bm{k}}{(2\pi)^3} \,  \frac{e^{i \bm{k} \cdot \bm{r}}}{k^2 + i 0^{^+}}  = 
- \frac{1}{4 \pi} \frac{e^{i {\left| \omega \right|} r}}{r}
\end{equation}
By substituting Eq.~\eqref{Massless Propagator} in Eq.~\eqref{Casimir - Polder Series} and differentiating 
\begin{eqnarray}
\label{eq:U-massless}
U {\left( \mu {=} 0{;}r \right)} {=}&&  \frac{4 i}{ (4 \pi)^3\,r^6} {\sum_{n {=} 0}^{\infty}} {\int_0^{\infty}}\, d\omega \, {g}_{0} {\left( {-} i \omega r \right)} { {\left( 2 \omega r \right)} }^{2 n} \times \nonumber \\&& e^{2 i \omega {\left( r {+} i 0^+ \right)} } \,  \left( \frac{1}{2 r} \right)^{2 n}\, \frac{1}{ \left( 2 n \right)!} \times \nonumber \\ &&  {\sum_{k {=} 0}^n} \binom{2 n}{2 k} \frac{d^{2 k} {\alpha}_1}{d {\omega}^{2 k}} {\left( 0 \right)} \frac{d^{2 n {-} 2 k} {\alpha}_2}{d {\omega}^{2 n {-} 2 k}} {\left( 0 \right)}
\end{eqnarray}
where 
\begin{equation}
{g}_{0} {\left( x \right)} = x^4 + 2 x^3 + 5 x^2 + 6 x + 3,
\end{equation}
and we have regularized the integral. 
\\
 Changing the variable $x = 2 \omega r$  in Eq.~\eqref{eq:U-massless}
\begin{eqnarray}
U {\left( \mu {=} 0{;}r \right)} {=}&& \frac{1}{{ {\left( 4 \pi \right)} }^3} \frac{2 i}{r^7} \sum_{n {=} 0}^{\infty} \int_0^{\infty}\, d x\, {g}_{0} {\left( {-} i \frac{x}{2} \right)} x^{2 n} e^{i x {\left( 1 {+} i 0^+ \right)} }  \nonumber \\ && \times \frac{{ {\left( \frac{1}{2 r} \right)} }^{2 n}}{ {\left( 2 n \right)} !} \sum_{k {=} 0}^n \binom{2 n}{2 k} \frac{d^{2 k} {\alpha}_1}{d {\omega}^{2 k}} {\left( 0 \right)} \frac{d^{2 n {-} 2 k} {\alpha}_2}{d {\omega}^{2 n {-} 2 k}} {\left( 0 \right)}\nonumber \\
\end{eqnarray}
and using the formula 
\begin{equation}
\int_0^{\infty} x^n e^{i x {\left( 1 {+} i 0^+ \right)} } d x = i^{n + 1} n!
\end{equation}
we get the final result:
\begin{eqnarray}\label{Massless Casimir - Polder Series}
U {\left( \mu {=} 0{;}r \right)} {=}&& {-} \frac{1}{{ {\left( 4 \pi \right)} }^3} \frac{1}{r^7} {\sum_{n {=} 0}^{\infty}} \frac{ {\left( 4 n^2 {+} 16 n {+} 23 \right)} {\left( n {+} 2 \right)} {\left( n {+} 1 \right)} }{2}\nonumber \\ &&\times{ {\left( {-} \frac{1}{4 r^2} \right)} }^n {\sum_{k {=} 0}^n} \binom{2 n}{2 k} \frac{d^{2 k} {\alpha}_1}{d {\omega}^{2 k}} {\left( 0 \right)} \frac{d^{2 n {-} 2 k} {\alpha}_2}{d {\omega}^{2 n {-} 2 k}} {\left( 0 \right)}.\nonumber\\\end{eqnarray}
The leading term of the series is the well known Casimir-Polder potential energy between two neutral atomic systems with static polarizability  $\alpha_{1,2}(0)$ given in Eq.~\eqref{Massless Casimir - Polder}. 

\subsection{Casimir - Polder interaction for massive photons} \label{submassive}
By using standard mathematical procedures (Jordan's lemma and Cauchy's residue theorem) we get from Eq.~\eqref{Dfunction} in the case $\mu \ne 0$:
\begin{equation}\label{Massive Propagator}
D {\left( \omega,r \right)} = - \frac{1}{4 \pi r} 
\begin{cases}
e^{- \sqrt{{\mu}^2 - {\omega}^2} r} & {\left( {\left| \omega \right|} \leq \mu \right)} \\
e^{i \sqrt{{\omega}^2 - {\mu}^2} r} & {\left( {\left| \omega \right|} \geq \mu \right)}
\end{cases}
\end{equation}
By substituting \eqref{Massive Propagator} in \eqref{Casimir - Polder Series}, differentiating and performing the appropriate change of variables   one obtains the following expression:
\begin{equation}
U {\left( \mu;r \right)} = U_1 {\left( \mu;r \right)} + U_2 {\left( \mu;r \right)}
\end{equation}
\begin{eqnarray}
U_1 \left( \mu;r \right) &=& - \frac{1}{{ {\left( 4 \pi \right)} }^3} \frac{1}{r^7}  \sum_{n=0}^{\infty} \int_0^{2 \mu r} \frac{xdx}{2}{ {\left[ { {\left( 2 \mu r \right)} }^2 - x^2 \right]} }^{n - \frac{1}{2}} \times  \nonumber \\ && {g}_{\mu} {\left( \frac{x}{2} \right)} e^{- x} \,  \left( \frac{1}{2 r} \right)^{2 n}\, \frac{1}{ \left( 2 n \right)!} \times \nonumber \\
&& \sum_{k=0}^n \binom{2 n}{2 k} \frac{d^{2 k} {\alpha}_1}{d {\omega}^{2 k}} {\left( 0 \right)} \frac{d^{2 n - 2 k} {\alpha}_2}{d {\omega}^{2 n - 2 k}} {\left( 0 \right)}
\label{integral1}
\end{eqnarray}
\begin{eqnarray}
U_2 {\left( \mu;r \right)} &=& - \frac{1}{{ {\left( 4 \pi \right)} }^3} \frac{1}{r^7}  \sum_{n=0}^{\infty} \int_0^{\infty} \frac{x dx }{2} { {\left[ { {\left( 2 \mu r \right)} }^2 + x^2 \right]} }^{n - \frac{1}{2}}   \nonumber \times \\ && {g}_{\mu} {\left( - i \frac{x}{2} \right)} e^{i x {\left( 1 + i 0^+ \right)} }\, \left( \frac{1}{2 r} \right)^{2 n}\, \frac{1}{ \left( 2 n \right)!} \times \nonumber\\
&& \sum_{k=0}^n \binom{2 n}{2 k} \frac{d^{2 k} {\alpha}_1}{d {\omega}^{2 k}} {\left( 0 \right)} \frac{d^{2 n - 2 k} {\alpha}_2}{d {\omega}^{2 n - 2 k}} {\left( 0 \right)},
\label{integral2}
\end{eqnarray}
where in this case (both in Eq.~\eqref{integral1} and Eq.~\eqref{integral2}): 
\begin{eqnarray}
&&{g}_{\mu} {\left( x \right)} =\nonumber \\ && \frac{2}{i} {\left[ 3 { {\left( \mu r \right)} }^4 - 4 { {\left( \mu r \right)} }^2 x^2 + 2 {\left( x^4 + 2 x^3 + 5 x^2 + 6 x + 3 \right)} \right]} \nonumber \\
\end{eqnarray}
and we have regularized the integral in Eq.~\eqref{integral2}.
Therefore, for the massive case, the final result reads 
\begin{eqnarray}\label{Massive Casimir - Polder Series}
 U {\left( \mu;r \right)} =&& - \frac{1}{{ {\left( 4 \pi \right)} }^3} \frac{1}{r^7} \sum_{n=0}^{\infty} \left[ 2 K_{n + 1} {\left( 2 \mu r \right)} { {\left( \mu r \right)} }^{n + 5} \right. \nonumber \\ 
 &&  {+} 8 K_{n + 2} {\left( 2 \mu r \right)} { {\left( \mu r \right)} }^{n + 4} \nonumber \\ &&\left. {+} {\left( 4 n^2 {+} 16 n {+} 23 \right)} K_{n + 3} {\left( 2 \mu r \right)} { {\left( \mu r \right)} }^{n + 3} \right] \nonumber\\ 
 &&\times \frac{{ {\left( - \frac{1}{4 r^2} \right)} }^n}{n!} 
 \sum_{k=0}^n \binom{2 n}{2 k} \frac{d^{2 k} {\alpha}_1}{d {\omega}^{2 k}} {\left( 0 \right)} \frac{d^{2 n - 2 k} {\alpha}_2}{d {\omega}^{2 n - 2 k}} {\left( 0 \right)}\nonumber\\
\end{eqnarray}
where $K_n {\left( x \right)}$ is the modified Bessel function of the second kind.  Performing the limit $\mu \rightarrow 0$: %
\begin{equation}
\lim_{x \to 0^{^+}} K_n {\left( 2 x \right)} x^n = \frac{ {\left( n - 1 \right)} !}{2}
\end{equation}
we see that in the massless limit of Eq.~\eqref{Massive Casimir - Polder Series} we recover the series in Eq.~\eqref{Massless Casimir - Polder Series}.
 The first (dominant) term in the above series \eqref{Massive Casimir - Polder Series} is:
\begin{eqnarray}\label{massiveCP-first-term}
 U \left( \mu \right.&;&\left. r \right) = - \frac{1}{\left( 4 \pi \right)^3} \frac{\alpha_1(0)\, \alpha_2(0)}{r^7} \left[ 2\,K_1(2\mu r) (\mu r)^5  \right. \nonumber\\&& +\left. 8\, K_2(2\mu r) (\mu r)^4 +23\, K_3(2\mu r)(\mu r)^3\right].
 \end{eqnarray}
 And of course in the limit $\mu \to 0$ from Eq.~\eqref{massiveCP-first-term} we readily recover the standard QED Casimir-Polder result, i.e. Eq.~\eqref{Massless Casimir - Polder}.
%
%
%
 We  emphasize that  Eq.~\eqref{Massive Casimir - Polder Series} and Eq.~\eqref{massiveCP-first-term} are the central results of the present work. In the following, we will use mainly Eq.~\eqref{massiveCP-first-term} (the leading term) to address some beyond the standard model scenarios with respect to the Casimir-Polder interactions.  

\section{Casimir-polder interaction in BSM models}
\label{sec:bsm}
In this section we discuss the Casimir-polder interactions in a number of alternative scenarios of physics beyond the standard model (BSM). Specifically,  we consider:  (A) universal extra dimensions,  (B) Randall Sundrum models and  (C) scale-invariant models. For each of the above BSM scenarios, the standard Casimir effect has been already discussed in the literature (see references in the introduction and in the following sections). 

\subsection{Universal Extra Dimensions (UED)}

\noindent In the UED scenario~\cite{Fabiano:2008aa}, all standard model fields are assumed to propagate in a bulk space-time with extra space-like dimensions compactified to a circle of radius $R$. Upon quantization of the (4+D) dimensional theory the effect of the extra dimension(s) is, for a given standard model field $X$, that there  is a tower of Kaluza-Klein states $X_n$, $n=1,2,\dots$ with masses:
\begin{equation}
m^2_n = m^2_{0}+\frac{n^2}{R^2}
\end{equation} 
where $m_0$ is the mass of the lowest lying state $X_0$. For the photon $m_0=0$ and   
thus the photon is then accompained by a Kaluza-Klein tower (KK-tower) of massive photons $\gamma_n$ of mass $m_n= n/R$.
Recent bounds on the size of the extra dimensions come from the non observation of Kaluza-Klein excitations at Tevatron and are already quite stringent: $R\le 300\,\, \text{GeV}^{-1}\approx 10^{-9}\,\, \text{nm}$~\cite{Appelquist:2001aa,Macesanu:2002aa,Rizzo:2001aa}. Stronger bounds are of course now available from LHC experiments and typical ATLAS~\cite{Aad:2015aa} and CMS a\-na\-ly\-ses exclude now values of $R^{-1}$ smaller than $\sim (1\, \text{TeV})^{-1}$~\cite{Beuria:2018aa} (or equivalently the allowed values of $R$ are those such that $R \le  \text{1\ TeV}^{-1} \approx 0.3 \times 10^{-9} \text{nm}$). Interestingly,  considerations from the relic density in the UED model assuming $\gamma_1$ to be the lightest Kaluza-Klein particle (LKP) give a preferred range for the size of the extra dimension: $R^{-1}\sim 1.3 -1.5$ TeV~\cite{Belanger:2011aa} thus  providing also an \emph{upper} bound for $R^{-1}$ (or a lower bound for $R$). 

The Casimir effect in the geometry of parallel conductor plates within UED has been addressed in~\cite{Poppenhaeger:2003es,2008BrJPh..38..581P}, and 
the bounds that can be obtained  are quite less stringent: $R< 10$ nm~\cite{Poppenhaeger:2003es}.
Given that typical current state of the art  Casimir/Casimir-Polder experiments can probe distances ($r$) down to the nanometer range if we assume the more stringent high energy bound on the compactification size $R$ of the extra dimension ($R\approx 0.3 \times 10^{-9}$ nm) the quantity $\xi=r/R$ is a very large quantity $\xi \approx 10^{10}$. Clearly for each one of the Kaluza-Klein massive photons $\gamma_n$ we can compute its contribution to the Casimir-Polder interaction of the two neutral systems via the result obtained in the previous section for the massive photon case in Eq.~\eqref{Massive Casimir - Polder Series}. In particular let us consider only the dominant term in the series in Eq.~\eqref{Massive Casimir - Polder Series}, that is the approximation in  Eq.~\eqref{massiveCP-first-term}. We can then estimate the total KK-tower contribution as summing, for every mass eigen-state, a term given by Eq.~\eqref{massiveCP-first-term}:
\begin{eqnarray}
\label{UKK}
U_{\text{KK}}&&\!\!\!\!\!\!\!(r, R) = \sum_{n=1}^{\infty} U\left(\frac{n}{R}, r\right) \nonumber \\
&=& - \frac{1}{(4\pi)^3} \frac{\alpha_1(0)\, \alpha_2(0)}{r^7} \sum_{n=1}^{\infty} \left[ 2 K_1\left(2n\xi\right) \left(n\xi\right)^5\right.\nonumber \\
&&+\left. 8 K_2\left(2n\xi\right) \left(n\xi\right)^4 +
23 K_3\left(2n\xi\right) \left(n\xi\right)^3\right]_{\xi={r}/{R}}\nonumber \\
\end{eqnarray}
It turns out that the above series converges very quickly and even a  truncation with a limited number of terms yields a stable numerical output. Therefore the expression of $U_{KK}$ in Eq~\eqref{UKK} can be easily computed numerically.   We can nevertheless provide a compact integral representation of Eq~\eqref{UKK} in terms of the PolyLog special functions which will be also useful for numerical estimates. We recall the following integral representation for the modified bessel functions~\cite{Abramowitz:1964aa}:
\[K_{\nu}(z)=\frac{\sqrt{\pi}(z / 2)^{\nu}}{\Gamma\left(\nu+\frac{1}{2}\right)} \int_{0}^{-\infty} e^{-z t}\left(t^{2}-1\right)^{\nu-1 / 2} d t
\]
which allows to rewrite the infinite sums appearing in Eq.~\eqref{UKK} as:
\begin{eqnarray}
\sum_{n=1}^{\infty} K_{\nu}(2 n \xi)(n \xi)^{\alpha}&=&
 \frac{\sqrt{\pi}}{\Gamma\left(\nu+\frac{1}{2}\right)}\int_{1}^{\infty}d t \left(t^{2}-1\right)^{\nu-\frac{1}{2}} \times\nonumber\\&&  \left[\sum_{n=1}^{\infty} (n \xi)^{\nu+\alpha} e^{-2 n \xi t}\right]
\end{eqnarray}
The series in the above relation can be rewritten in terms of a $\text{PolyLog}$ function: $\text{PolyLog}[\alpha, z]= \sum_{n=1}^\infty z^n/n^\alpha$:
\begin{eqnarray}
\sum_{n=1}^{\infty} K_{\nu}(2 n \xi)(n \xi)^{\alpha}&=&
 \frac{\sqrt{\pi}\xi^{\nu+\alpha}}{\Gamma\left(\nu+\frac{1}{2}\right)}\, \int_{1}^{\infty} d t\left(t^{2}-1\right)^{\nu-\frac{1}{2}} \nonumber \times \\ &&\text{PolyLog}[-\nu-\alpha,e^{-2\xi t}] 
\end{eqnarray}
Using the above result in Eq.~\eqref{UKK} we can finally give an exact integral representation of the ratio of the $U_{KK}$ potential to the QED potential $U$ of  massless Casimir-Polder, Eq.~\eqref{Massless Casimir - Polder} in terms of a PolyLog funtion:
\begin{eqnarray}
\label{UKKintegral}
    \frac{U_{KK}}{U}&=& \frac{\sqrt{\pi} \xi^{6}}{23 \, \Gamma(3 / 2)}
    \int_{1}^{\infty} \,d x\,  \text{PolyLog} \left[-6, e^{-2 \xi t}\right]\times \nonumber \\
    && \sqrt{t^{2}-1}\left[2+\frac{16}{3}\left(t^{2}-1\right)+\frac{92}{15}\left(t^{2}-1\right)^{2}\right]_{\xi=\frac{r}{R}}\nonumber\\
\end{eqnarray}
which may be used for a fast numerical 
computation of the effect. 

Further we  discuss 
an approximation of the above result or Eq.~\eqref{UKK} given by a finite number of terms. 

Given the constraints on the extra-dimension length $R$ we can assume  the relevant values of the ratio will be such that $r \gg R$ or $r/R \gg 1$ and  we can also give an approximated formula given by a finite number of terms which might be useful for practical purposes. 
In order to get the approximate formula we note that PolyLog functions of negative order satisfy the following property:
\begin{equation}
\label{negativePolyLog}
\begin{aligned} \sum_{k=1}^{\infty} k^{n} z^{k} &=\text{PolyLog}[-n,z]
\\ &=\frac{1}{(1-z)^{n+1}} \sum_{i=0}^{n-1} \left\langle \begin{array}{c}{n} \\ {i}\end{array}\right\rangle z^{n-i} \end{aligned}
\end{equation}
i.e. they reduce to a finite number of terms where the quantities $\left \langle n \atop i \right \rangle$ are the Eulerian number or the number of permutations of the numbers from 1 to $n$ in which exactly $i$ elements are greater than the previous element (permutations with $i$ "ascents") -- they are the coefficients of the Eulerian polynomials--.
In turn the approximation in Eq.~\eqref{UKKintegral} consists in assuming that the radial distance $r$ is large enough, so that $\xi=r/R \gg 1$,  and  the prefactor in right hand side of Eq.~\eqref{negativePolyLog} can be approximated by 1: $z= e^{-2\xi t}\ll 1$ and $(1- z)^{-7} \approx 1$.  Then the integrals in the $t$ variable in Eq.~\eqref{UKKintegral} can be computed analytically and the final result for $U_{KK}/U$ is:
\begin{equation}
    \label{UKK_approximated}
\begin{aligned}
\frac{U_{KK}}{U}=\sum_{k=0}^{5}\left\langle \begin{array}{c}{6} \\ {k}\end{array}\right\rangle\Bigg[&2  \frac{K_{1}\left(2(6-k) \xi\right)}{6-k}\,\xi^{5}+\\
&8 \ \frac{K_{2}\left(2(6-k) \xi\right)}{(6-k)^{2}}\,\xi^{4}+\\
&23  \frac{K_{3}\left(2(6-k)\xi\right)}{(6-k)^{3}}\,\xi^{3}\Bigg]_{\xi=\frac{r}{R}}
    \end{aligned}    
\end{equation}


\subsection{Randall-Sundrum models}
 We recall here  that in the Randall-Sundrum (RS) model~\cite{Randall:1999aa,PhysRevLett.83.4690,Teo2010} the underlying spacetime   is a  5D  anti-deSitter space $\left(\mathrm{AdS}_{5}\right)$ with background metric: 
\begin{equation}
\label{metricAdS5}
d s^{2}=g_{a b} d x^{a} d x^{b}=e^{-2 \kappa|y|} \eta_{\mu\nu } d x^{\mu} d x^{\nu}-d y^{2}
\end{equation}
 where  $\eta_{\mu \nu}=\operatorname{diag}(1,-1,-1,-1)$  is the usual  four dimensional (4D) Minkowski spacetime metric. 
 In Eq.~\eqref{metricAdS5} $\mu, \nu $ stand for  the  indices of the $4D$ 3-brane and they assume the usual values  from  0 to  3,  while  $(a, b)$ are the indices in the  5D bulk ranging  from 0 to  4.
 The  $y$ coordinate describes the space-like extra dimension which is  compactified   on the orbifold $S^{1} / \mathbb{Z}_{2}$.
 We see that the ($4D$) Minkowski metric is multiplied by a  factor, $e^{-2 \kappa|y|}$, which depends on the coordinate of the extra dimension $y$ through the parameter
 $\kappa$  in terms of which  is expressed 
 the curvature tensor of the underlying  space (AdS$_5$). The model is characterized by a visible 3-brane at $y=0$ and an invisible one at $y=\pi R$ with opposite tension, $R$ being the compactification radius of the extra spacelike dimension described by the coordinate $y$.
The  mass spectrum of the Randall-Sundrum model is characterized by a tower of Kaluza-Klein states which, differently from the UED model, are exponentially suppressed. For the scalar field the KK tower is given by:
\begin{equation}
\label{RS_spectrum_scalar}
\frac{m_N\, e^{\kappa R}}{\kappa} \approx \pi \left( N + \frac{1}{4}\right)\qquad \qquad N=1,2,3 \dots
\end{equation}
The standard Casimir effect between conducting parallel plates in Randall-Sundrum models has been considered first in~\cite{Frank:2007ab} by adopting the scalar field analogy. However here in order to estimate  the Casimir-Polder interaction via Eq.~\eqref{Massive Casimir - Polder Series} we need to consider the KK spectrum of a vector field which is different from that of a scalar field in Eq.~\eqref{RS_spectrum_scalar}, and is given by~\cite{Teo2010}:
\begin{equation}
m_N=\kappa\, z_N \qquad \qquad N=1,2,3 \dots
\end{equation}
where $z_N$ are the roots, in the $z$ variable, of the equation:
\begin{equation}
\label{RS_spectrum}
J _ { 0 } \left( { z }  \right) Y _ { 0 } \left(  { z e ^ { \pi \kappa R } }  \right) - Y _ { 0 } \left( { z }  \right) J _ { 0 } \left(  { z e ^ { \pi \kappa R } }  \right) = 0\, . 
\end{equation}
The approximated roots of Eq.~\eqref{RS_spectrum} are
\begin{equation}
\label{RS_spectrum_scalar_1}
z_N \approx N \frac{ \pi }{ { e }^{ \pi \kappa R} - 1 }\qquad \qquad N = 1,2,3 \dots
\end{equation}
We can then compute the mass spectrum over which we will have to sum Eq.~\eqref{Massive Casimir - Polder Series} and/or Eq.~\eqref{massiveCP-first-term} in order to get the Randall-Sundrum contribution to the Casimir-Polder potential. 
We have to compute the quantity $\mu r$ into $m_N r$ so that using the mass spectrum 
Eq.~\eqref{RS_spectrum_scalar_1} we have:
\begin{equation}
\label{RS_spectrum_vector}
\begin{aligned}
    m_N r  &\approx N \frac{ \pi \kappa}{ { e }^{ \pi \kappa R} - 1 }r \\&= N \frac{ \pi \kappa R}{ { e }^{ \pi \kappa R} - 1 }\frac{r}{R}\qquad N = 1,2,3 \dots
\end{aligned}
\end{equation}
and if we set:
\begin{subequations}
\label{RS_parameters}
\begin{align}
	\label{RS_parameters_a}
a &= \frac{ \pi \kappa R }{ { e }^{ \pi \kappa R } - 1 }\\
\label{RS_parameters_b}
\xi &=\frac{r}{R}
\end{align}
\end{subequations}
we will have:
\begin{equation}
m_N r= N a \xi     \qquad \qquad N=1,2,3 \dots
\end{equation}
We conclude that the Randall-Sundrum Casimir-Polder effect will be given again by the same formulas obtained for the UED model, Eqs.~(\ref{UKK},~\ref{UKKintegral},~\ref{UKK_approximated}), and  simply making there the replacement $\xi \to a \xi$, according to Eq.~\eqref{RS_spectrum_vector} and
Eqs.~\eqref{RS_parameters}.

\subsection{Unparticle Casimir-Polder}

Based on the conjecture~\cite{PhysRevLett.98.221601}, we examine now a model that introduces a new massive sector in the SM able to preserve scale invariance properties.
However, this is valid only under the condition of exhibiting a non-integer number of particles $d_{\mathcal{U}}$.
In particular, for massive fields, scale invariance can be described by the so called Banks-Zacks  fields ($\cal{BZ}$) ~\cite{Banks:1981nn}. Then, in the unparticle description, one can say that there is an energy scale $\Lambda_{\mathcal{U}}$ that sets the transition between free particle behaviour at high energies and unparticle behaviour at lower energies.
At this energy scale $\Lambda_{\mathcal{U}}$, the $\cal{BZ}$ sector shows scale-invariant properties and the number of particles is controlled by
$d_{\mathcal{U}}$. This parameter is generally restricted to be $2 \ge d_{\mathcal{U}} \ge 1$. 
Where the lower bound is given by unitarity constraints from conformal field theory (CFT)~\cite{Grinstein2008367} while the higher bound $d_{\mathcal{U}} \ge 2$ is introduced because the calculations are less predictive due to the ultraviolet sector.

In a recent work~\cite{Frassino:2017aa, Frassino:2019aa} some of the present authors derived the Casimir effect for the unparticle field in the geometry of the parallel conductor plates. 
The central result is that the unparticle Casimir energy is given by a mass  integral over the Casimir energy at given mass:
\begin{equation}
\label{Uncasimir}
\mathcal { E } _ {\cal U } ^ { \mathrm { C } } = \frac { A _ { d_{\cal U} } } { \pi \left( \Lambda _ { \mathcal { U } } ^ { 2 } \right) ^ { d _ { \mathcal { U } } - 1 } } \int _ { 0 } ^ { \infty } d \mu \mu ^ { 2 d _ { \mathcal { U } } - 3 } \mathcal { E } ^ { \mathrm { C } } ( \mu )
\end{equation}
where $A_{d_{\cal U}}$ is a numerical constant: 
\begin{equation}
A_{d_{\cal U}} \equiv \frac{16 \pi^{5 / 2}}{(2 \pi)^{2 d_{\cal U}}} \frac{\Gamma\left(d_{\cal U}+1 / 2\right)}{\Gamma\left(d_{\cal U}-1\right) \Gamma\left(2 d_{U}\right)}
\, ,
\end{equation}
routinely used in the literature of unaprticle phenomenology.
But it is well known that the Casimir energy 
${\cal E}^C$ 
between two parallel plates (and in general between two given surfaces of arbitrary geometrical shape)  ~\cite{Milton_Book,Farina:2006hk}
can be related
to a pairwise integration of the  Casimir-Polder interaction $U({\bm{r}_1}-{\bm{r}_2})$:
\begin{equation}
\label{Casimir-CP}
 {\cal E}^C = \frac{{\cal N}^2}{2}\int_0^a dz_1 \int_0^a dz_2
 \int d^2{\bm{r}_1}_\perp\int d^2{\bm{r}_2}_\perp\,
 U ({\bm{r}_1}-{\bm{r}_2}) \,    
\end{equation} 
where ${\cal N}$ is the number of atom/molecules per unit volume (number density) in the conductor plates (surfaces)~\cite{Milton:1978sf,Milton:1997ky}. The above result can be extended straight-forwardly  to the massive case.
A relation similar to Eq.~\eqref{Casimir-CP} is also expected to hold between the unparticle casimir energy $ {\cal E}^C_{\cal U}$ between perfect conductor plates  and  the unparticle Casimir-Polder interactions $U_{d_{\cal U}}({\bm{r}_1}-{\bm{r}_2})$ between atomic and/or molecular systems: 
\begin{eqnarray}
\label{Uncasimir-UCP}
 {\cal E}^C_{\cal U} =&&\frac{{\cal N}^2}{2}\nonumber \times\\&&\int_0^a dz_1 \int_0^a dz_2
 \int d^2{\bm{r}_1}_\perp\int d^2{\bm{r}_2}_\perp\,  U_{d_{\cal U}}({\bm{r}_1}-{\bm{r}_2})\nonumber\\    
\end{eqnarray}
Then by inserting Eq.~\eqref{Uncasimir-UCP} and Eq.~\eqref{Casimir-CP} respectively in the left and right members of Eq.~\eqref{Uncasimir} and 
given the arbitrarity of the geometry considered we can infer that  the unparticle Casimir-Polder potential energy, $U_{d_{\cal U}}(r)$ with  $r=|{\bm{r}_1}-{\bm{r}_2}|$,   is the superposition of Casimir-Polder interactions at finite mass $\mu$, $U(\mu,r)$,
and  we have:
\begin{equation}\label{Integral}
U_{ d_{\cal U}}(r )=  \frac{A_{d_{\cal U}}}{\pi({{\Lambda}_{\cal U}^2})^{ d_{\cal U} - 1}}\, \int_0^{\infty}  \, d \mu\,{\mu}^{2 d_{\cal U} - 3}\,U {\left( \mu;r \right)} 
\end{equation}
By substituting Eq.~\eqref{Massive Casimir - Polder Series} into Eq.~\eqref{Integral} and using the formula~\cite{Olver:2010aa}: 
\begin{equation}
\int_0^{\infty} K_a {\left( 2 x \right)} x^b d x = \frac{1}{4} \Gamma {\left( \frac{a + b + 1}{2} \right)} \Gamma {\left( \frac{b - a + 1}{2} \right)}
\end{equation}
we get:
\begin{widetext}
\begin{equation}\label{Unparticle Casimir - Polder Series}
\begin{split}
 U_{ d_{\cal U}}(r ) &= - \frac{1}{{ {\left( 4 \pi \right)} }^{2 d_{\cal U} + 1}} \frac{1}{r^7 { {\left( {\Lambda}_{\cal U} r \right)} }^{2 d_{\cal U} - 2}} \times \\
&\phantom{=} \sum_{n=0}^{\infty} {\left[ {2 d_{\cal U}}^2 + 6 d_{\cal U} + {\left( 2 n + 3 \right)} {\left( 2 n + 5 \right)} \right]} \frac{\Gamma {\left( n + d_{\cal U} + 2 \right)} }{2 \Gamma {\left( n + 1 \right)} \Gamma {\left( d_{\cal U} \right)} } { {\left( - \frac{1}{4 r^2} \right)} }^n \sum_{k=0}^n \binom{2 n}{2 k} \frac{d^{2 k} {\alpha}^{\scriptscriptstyle \cal U}_1}{d {\omega}^{2 k}} {\left( 0 \right)} \frac{d^{2 n - 2 k} {\alpha}^{\scriptscriptstyle \cal U}_2}{d {\omega}^{2 n - 2 k}} {\left( 0 \right)}
\end{split}
\end{equation}
In the ``particle limit'' ($d_{\cal U} \to 1$) of Eq.~(\ref{Unparticle Casimir - Polder Series}) we recover the series in Eq.~(\ref{Massless Casimir - Polder Series}). 
%
\end{widetext}
The leading term of the series \eqref{Unparticle Casimir - Polder Series} is the potential energy \\
\begin{eqnarray}
U_{d_{\cal U}}(r) &=& - \frac{1}{{ {\left( 4 \pi \right)} }^{2 d_{\cal U} + 1}} \frac{{\alpha}^{\scriptscriptstyle \cal U}_1 {\left( 0 \right)} {\alpha}^{\scriptscriptstyle \cal U}_2 {\left( 0 \right)} }{r^7 { {\left( {\Lambda}_{\cal U} r \right)} }^{2 d_{\cal U} - 2}}\,\times\nonumber\\ &&\frac{ {\left( {2 d_{\cal U}}^2 + 6 d_{\cal U} + 15 \right)} {\left( d_{\cal U} + 1 \right)} d_{\cal U}}{2}
\end{eqnarray}
If we assume that the unparticle charges of protons and electrons are opposite and equal in absolute value to $\lambda$, atoms and molecules are neutral also with respect to the unparticle charge, and non - polar molecules are non - polar with respect to the Unparticle charge too. 
Therefore the Unparticle dipole is:
\begin{equation*}
d_{\, \scriptstyle \cal U} = \frac{\lambda}{e}\, d
\end{equation*}
and by using \eqref{Braket Dipole} the Unparticle polarizability is
\begin{equation*}
{\alpha}^{\,\scriptstyle \cal U} (\omega)= { {\left( \frac{\lambda}{e} \right)} }^2 \alpha(\omega)
\end{equation*}
Therefore the final result is 
\begin{eqnarray}\label{Unparticle Casimir - Polder}
U_{ d_{\cal U}}(r) &=& - \frac{1}{{ {\left( 4 \pi \right)} }^{2 d_{\cal U} + 1}} { {\left( \frac{\lambda}{e} \right)} }^4 \frac{{\alpha}_1 {\left( 0 \right)} {\alpha}_2 {\left( 0 \right)} }{r^7 { {\left( {\Lambda}_{\cal U} r \right)} }^{2 d_{\cal U} - 2}}\times \nonumber \\ && \frac{ {\left( {2 d_{\cal U}}^2 + 6 d_{\cal U} + 15 \right)} {\left( d_{\cal U} + 1 \right)} d_{\cal U}}{2}\,.
\end{eqnarray}
{The ratio of the unparticle contribution $U_{d_{\cal U}}$ to the standard QED result $U$, Eq.~\eqref{Massless Casimir - Polder}, is therefore:}
\begin{equation}\label{ratio_unparticle}
\frac{U_{d_{\cal U}}}{U}  =  { {\left( \frac{\lambda}{e} \right)} }^4 \,  \frac{ {\left( {2 d_{\cal U}}^2 + 6 d_{\cal U} + 15 \right)} {\left( d_{\cal U} + 1 \right)} d_{\cal U}}{2\times 23\, { {\left( 4 \pi \right)} }^{2 d_{\cal U} - 2} { { {\left( {\Lambda}_{\cal U} r \right)} }^{2 d_{\cal U} - 2}} }\,.
\end{equation}

\section{Discussion and results}
We now discuss the previous analytical results and provide some numerical estimates of the Casimir-Polder contribution of the various BSM models considered in the previous Section~\ref{sec:bsm} relative to the  Casimir-Polder in standard quantum electro-dynamics (QED).\\

\subsection{Universal Extra Dimensions}

\begin{figure}[t!]
	\begin{centering}
	\includegraphics[scale=0.55]{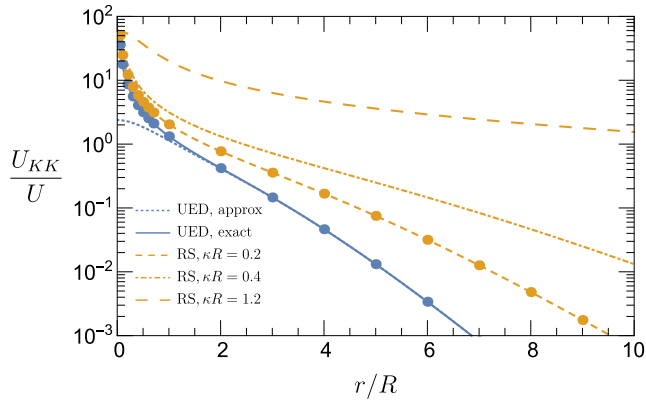}
	\end{centering}
	\caption{\label{fig:UED}Contribution of the  KK-tower in the universal extra dimension model to the Casimir-Polder interaction potential relative to the standard QED (massless photon). The solid (blue) line is the  numerical result from  Eq.~\protect\eqref{UKK} obtained truncating the series at $N_{\text{max}}=60$. The full (blue) disks are the results from Eq.~\protect\eqref{UKKintegral}. The dotted (blue) line is the approximated form as given in Eq.~\protect\eqref{UKK_approximated}. The other curves represent the Randall-Sundrum contribution for diffeerent values of the dimensionless parameter $\kappa R $ (with $\xi \to a \xi$ and $a$ given in Eq.~\ref{RS_parameters_a}) respectivley equal to $ 0.2$ dashed (orange), $0.4$ dot-dashed (orange) and $1.2$ long-dashed (orange). The full (orange) disks superimposed with the dashed curve represent the result through Eq.~\ref{UKKintegral} with $\xi \to a \xi$. }
\end{figure}

In Fig.~\ref{fig:UED} we show the contribution of the KK tower of massive states both in universal extra dimensions (UED) and in Randall-Sundrum (RS) models   $U_{KK}$ relative to the standard QED result (for a massless photon).
If one assumes the current bound~\cite{Choudhury:2016aa} from direct searches at particle accelerators (LHC) $R\le (1\,\, \text{TeV})^{-1}\approx 0.3 \times 10^{-9}\,\, \text{nm}$ then the deviations to the Casimir-Polder interaction from the Kaluza-Klein tower would be entirely negligible since distances $r$ that can be probed in current state of the art Casimir/Casimir-Polder experiments~\cite{Klimchitskaya:2009aa} are at least in the nano-meter range (or larger)  then $\xi=r/R > 3.3\times 10^9$ and from Fig.~\ref{fig:UED} we see that the ratio ${U_{KK}}/{U}$ is ${\cal O}(10^{-3})$ already for $\xi\approx 10$ and decreases exponentially fast. One can see from Fig.~\ref{fig:UED} that the approximation in Eq.~\eqref{UKK_approximated} is quite good for values of the parameter $\xi=r/R \approx 1 $ or greater. 

{However the fact that typical distances in Casimir and Casimir-Polder experiments  range from the nanometer up to a few microns ($ 3.3 \times 10^9 \le  r/R \le 3.3 \times 10^{12})$ leaves little hope that within the UED model the Casimir-Polder interactions might actually be ever measured. From Fig.~\ref{fig:UED} it is clear that such high values of $r/R$ will provide an extremely small correction to the standard QED Casimir Polder.} 
\\

\subsection{Randall-Sundrum}

In Fig.~\ref{fig:UED} we also show  the contribution of the Randall-Sundrum KK tower relative to the standard Casimir-Polder $U_{KK}/U$ as a function of $\xi=r/R$. As dicussed above the  Casimir-Polder interaction in the Randall-Sundrum model is given by the same formulae of the UED case, Eqs.~(\ref{UKK},\ref{UKKintegral},\ref{UKK_approximated}), with the replacement $\xi \to a\xi$ with $a$ given in Eq.~\ref{RS_parameters_a}. We show the results for three different values of the dimensionless parameter $\kappa R=0.2,0.4,1.2$.
{From Fig.~\ref{fig:UED} we see that the Randall-Sundrum contribution to the Casimir-Polder interaction has a better chance of being non-negligible at  values of the distance of experimental interest (nanometers) for larger values of the parameter $\kappa R$. Indeed we find for instance that for a value of $\kappa R=8.2$ and  $R\approx 1$ TeV$^{-1} = 3\times 10^{9} \text{nm}^{-1}$ the ratio $U_{KK}/U$ is for $r=10$ nm ($\xi=3.3 \times 10^{10}$) about 0.04, or a 4\% contribution from the RS model.}
\\

\begin{figure}[!ht]
\begin{centering}
\includegraphics[scale=0.55]{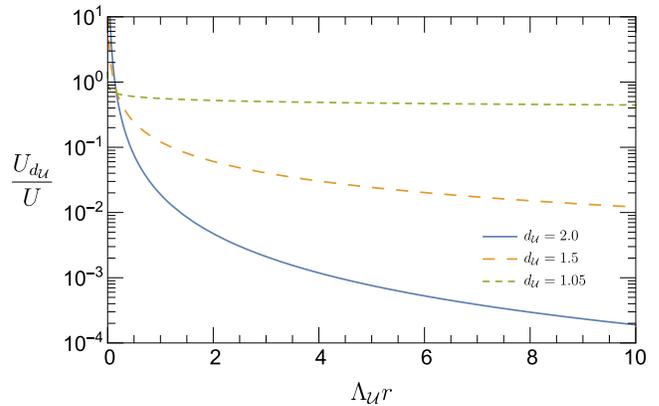}
\end{centering}
\caption{Unparticle  Casimir-Polder interaction potential $U_{d_{\cal U}}$ relative to the standard QED (massless photon) Casimir-Polder interaction $U$ as a function of the distance, in units of $\Lambda_{\cal U}^{-1}$, at a fixed value of the coupling $\lambda/e=0.9$. The solid (blue) line is the result for $d_{\cal U}=2$, the long-dashed (orange) line is for $d_{\cal U}=1.5$ result and the short-dashed (green) line is for $d_{\cal U}=1.05$. }
\label{fig:Unparticle}
\end{figure}

\subsection{Unparticles}

In Fig.~\ref{fig:Unparticle} we show the ratio of the Unparticle Casimir-Polder interaction, $U_{d_{\cal U}}$, to the standard massless photon QED Casimir-Polder potential, $U$,  versus $\Lambda_{\cal U}r$, or the distance in units of $\Lambda_{\cal U}^{-1}$ for different values of the scaling dimension of the Unparticle field $d_{\cal U}=1.05, 1.5, 2$.

{From Fig.~\ref{fig:Unparticle} we see that higher values of $d_{\cal U}$ have a better chance of providing a contribution to the Casimir-Polder interaction which has the potential of being measurable. Indeed  if we assume a scale of the unparticle model of the order of the TeV, $\Lambda_{\cal U} = 1$ TeV ($\approx 3 \times 10^{9}$ nm$^{-1}$),  with $d_{\cal U}= 1.05\,\,  (\text{and\ } \lambda=0.9)$, we obtain numerically that $U_{d_{\cal U}}/U=0.049$ for typical distances of Casimir experiments in the nano-meter range ($\xi=\Lambda_{\cal U}r \approx 3.3\times 10^{10}$), i.e. a 5\% contribution. The fact that the unparticle contribution becomes relevant and possibly detectable only for $d_{\cal U}$ values very close to unity parallels what has been found in the analysis of the Unparticle Casimir effect in ref.~\cite{Frassino:2017aa}.}

\section{\label{sec:conclusions}Conclusions }
The quite vast current literature of Casimir interactions in relation to a massive photon discusses  only the standard Casimir effect in the geometry of  two parallel conductor plates.
This had been studied in the pioneering work of Barton and Dombey~\cite{Barton:1984kx,Barton:1985aa} and subsequently taken up by several authors in other BSM scenarios~\cite{Frassino:2011aa,Blasone:2019aa,Frank:2007ab,Frank:2008aa,Teo_5,Poppenhaeger:2003es,Frassino:2017aa,Frassino:2019aa,Buoninfante:2019aa,Lambiase:2017aa,Blasone:2018nfy,Teo:2010ab,Edery:2008kd}, but always considering the Casimir effect between parallel plates. 

In this paper we have studied the intermolecular Casimir-Polder forces between neutral systems at a distance $r$ from each other, mediated by a massive vector field (assuming an electromagnetic-type coupling) thereby filling a gap in the existing literature of Casimir interactions.  

Although this might appear at first to be a computation with only a speculative 
interest it is instead of direct application in deriving the Casimir-Polder interactions between neutral systems in theories beyond the standard model, such as universal extra dimensions (UED), Randall-Sundrum (RS) models and scale invariant theories (Unparticles). Moreover we have discussed the impact of the contributions to the Casimir-Polder interactions in these BSM models relative to the QED contribution and our results could be used to discuss complementary bounds on those BSM theories with future experiments.
The above mentioned scenarios are on the other end receiving a lot of attention in other research domains and especially so they are  well studied at high energy colliders like the LHC or its future upgrades like the high luminosity or the high energy LHC (HL-LHC or HE-LHC)~\cite{Cid-Vidal:2018aa}, thus highlighting the complementary value of the present work. 
Specifically we have discussed, within the above BSM scenarios, the deviations of the Casimir-Polder interactions relative to the QED (massless photon) case as a function of the distance $r$ between the neutral systems and in relation to the model free parameters. While the  UED  contribution to the Casimir-Polder interaction appears to be too small to be measurable in current experiments we have found that both for the RS and the unparticle models there are values of the parameters for which a sizeable contribution would result which could, in principle, be detected. 

It is the author's opinion that even in the absence of observed deviations from the standard QED massless Casimir-Polder one could use the  results of the present work to propose bounds on the BSM theories parameter space (at least for the RS and Unparticle cases) that could be compared to those derived from other searches such as the standard Casimir effect and/or even high energy accelerator searches.

\section*{Acknowledgments} 
L.~M. wishes to acknowledge support from the Erasmus Traineeship Program which allowed a visit to the Frankfurt Institute for Advanced Studies (FIAS) where the early stages of this work were carried out under the supervision of Piero Nicolini. 
The work of L.~M. is currently supported by the National Science Centre of Poland under the Grant: {\scshape Sonata Bis}, No. 2015/18/E/ST1/00200. A.M.F. is supported by ERC Advanced Grant GravBHs-692951 and MEC grant FPA2016-76005-C2-2-P.

%

\end{document}